\documentclass[sigconf]{acmart}

\AtBeginDocument{%
  \providecommand\BibTeX{{
    \normalfont B\kern-0.5em{\scshape i\kern-0.25em b}\kern-0.8em\TeX}}}

\renewcommand\footnotetextcopyrightpermission[1]{}

\usepackage[T1]{fontenc}
\usepackage{xspace}
\usepackage{mdwlist}
\usepackage{graphicx}
\usepackage{enumitem}
\usepackage{amsmath}

\usepackage{subfigure}
\newcounter{glossy_enum}
\makecompactlist{list*}{list}
\newenvironment{glossy_enumerate}
{\begin{list*}[{\arabic{glossy_enum})}{\usecounter{glossy_enum} \topsep=0.2em \leftmargin=1.4em \itemindent=-0.0em}]}
{\end{list*}\vspace*{0.5em}}

\setenumerate{topsep=0.2em,leftmargin=2.4em,itemindent=-0.0em}

\newcommand{\capt}[1]{\mdseries{\emph{#1}}}
\newcommand{\code}[1]{\textbf{{\texttt{#1}}}}
\newcommand{\fakepar}[1]{\vspace{.5mm}\noindent\textbf{#1.}}
\newcommand\figref[1]{Fig.\,\ref{#1}}
\newcommand\secref[1]{Sec.\,\ref{#1}}

\newcommand{\vect}[1]{\ensuremath{\boldsymbol{#1}}}
\renewcommand\eqref[1]{Eq.\,\ref{#1}}

\newcommand\uf{$\mu$F\xspace}

\newcommand\hz{Hz\xspace}
\newcommand\secs{sec\xspace}
\newcommand\mhz{MHz\xspace}

\newcommand\kb{Kb\xspace}

\newcommand\smart{\textsc{Smart}\xspace}
\newcommand\greedy{\textsc{Greedy}\xspace}

\begin{document}

\title{The Case for Approximate Intermittent Computing}

\author[Bambusi et al.]{Fulvio Bambusi$^{*}$, Francesco Cerizzi$^{*}$, Yamin Lee$^{*}$, and Luca Mottola$^{*\dagger+}$}
\affiliation{$^*$Politecnico di Milano (Italy), $^\dagger$RI.SE Sweden, $^+$Uppsala University (Sweden),}

\begin{abstract}
  We present the concept of \emph{approximate intermittent computing} and concretely demonstrate its application.
  Intermittent computations stem from the erratic energy patterns caused by energy harvesting: computations unpredictably terminate whenever energy is insufficient and the application state is lost.
  Existing solutions maintain equivalence to continuous executions by creating persistent state on non-volatile memory, enabling \emph{stateful computations to cross power failures}.
  The performance penalty is massive: system throughput reduces while energy consumption increases.  
  In contrast, approximate intermittent computations trade the accuracy of the results for sparing \emph{the entire overhead} to maintain equivalence to a continuous execution.
  This is possible as we use approximation to \emph{limit the extent of stateful computations to the single power cycle}, enabling the system to \emph{completely} shift the energy budget for managing persistent state to useful computations towards an \emph{immediate} approximate result.
  To this end, we effectively reverse the regular formulation of approximate computing problems. 
  First, we apply approximate intermittent computing to human activity recognition.
  We design an anytime variation of support vector machines able to improve the accuracy of the classification as energy is available.
  We build a hw/sw prototype using kinetic energy and show a $7x$ improvement in system throughput compared to state-of-the-art system support for intermittent computing, while retaining 83\% accuracy in a setting where the best attainable accuracy is 88\%. 
  Next, we apply approximate intermittent computing in a sharply different scenario, that is, embedded image processing, using loop perforation.
  Using a different hw/sw prototype we build and diverse energy traces, we show a $5x$ improvement in system throughput compared to state-of-the-art system support for intermittent computing, while providing an equivalent output in 84\% of the cases.
\end{abstract}


\maketitle

\section{Introduction}

Ambient energy harvesting enables battery-less embedded sensing~\cite{batteryless-future,soil-termoelectric,bridge-sensor, traffic-flow-sensor,sensys20deployment,flicker,permadaq}.
However, energy from the environment is generally erratic, causing frequent and unanticipated power failures.
For example, harvesting energy from RF transmissions to execute a simple CRC calculation may lead to $16$ power failures over a $6$ sec period~\cite{harvesting-survey}.
Executions thus become \textit{intermittent}, as they consist of intervals of active computation interleaved by possibly long periods of recharging energy buffers~\cite{harvesting-survey}.

\fakepar{Prior art} Due to resource constraints, power failures normally cause a device to lose the application state.
To ensure forward progress across power failures, a variety of techniques exists that make use of \emph{persistent state} stored on non-volatile memory (NVM), as we elaborate in~\secref{sec:related}.
Persistent state is then reloaded from NVM when energy is back, so executions resume closer to the point of power outage rather than performing a complete reboot.

Most existing solutions~\cite{mementos,Hibernus,Hibernus++,chinchilla,HarvOS,DICE,ratchet, dino,alpaca,chain,coala,Ink} aim to make intermittent executions equivalent to their continuous counterparts, as we articulate in \secref{sec:related}.
Given the same inputs, for example, certain sensor readings, the results of intermittent executions must be exactly the same as those of a continuous one.
To achieve this, existing solutions employ persistent state to allow \emph{stateful computations} to cross power failures.
The price for this is enormous, in both \emph{energy} consumption  and latency until a result is available, and thus system \emph{throughput}.
The energy overhead may reach up to 350\% the cost of the application processing, mainly due to the use of energy-hungry NVM technology~\cite{ratchet}.
Depending on energy patterns and the time for energy buffers to recharge, a 10~ms processing in a continuous execution may take minutes in an intermittent one~\cite{batteryless-future}, reducing throughput.

Nonetheless, using persistent state to allow stateful computations to cross power failures has further implications.
In mixed-volatile platforms~\cite{msp430fr5969}, slices of main memory are mapped to NVM, for example, with FRAM technology.
As a result, intermittence anomalies~\cite{maioli21ewsn,brokenTM} appear due to re-execution of non-idempotent code, requiring additional time and energy to be corrected.
When using FRAM, wait cycles may be necessary to synchronize read/write operations with the microcontroller, further increasing energy overhead~\cite{msp430fr5969}. 
Fianlly, if the state of computations is to be retained across power failures, the state of peripherals must also be correspondingly persistent~\cite{sytare,restop,samoyed,karma}.
This increases the size of the persistent state, as it must include information related to peripheral states that may not be reflected in the systems' main memory, adding to the energy overhead.

\fakepar{Approximate intermittent computing} Our work starts from the observation that a number of embedded sensing applications, as in smart health~\cite{solanas2014smart}, ambient intelligence~\cite{cook2009ambient}, and environment monitoring~\cite{othman2012wireless}, expose wto specific characteristics:
\begin{glossy_enumerate}
\item Latency to obtain a result is key; a fitness tracker must process input signals as rapidly as possible, because the outcome might require immediate reactions. The longer it takes for a sample to be processed, the lower is the value of the analysis. 
\item Newer inputs are more important than older ones. A fitness tracker should process the most recent samples first, as they are representative of the current situation. Any further processing may merely represent a ``best-effort'' task.
\end{glossy_enumerate}
Most importantly, \emph{approximate results} are often tolerated.
This stems from the nature of data processing in these applications, including computer vision, machine learning, signal processing, and pattern identification.
These algorithms offer probabilistic guarantees in the first place and are robust to data errors, for example, due to sensor inaccuracies~\cite{mittal2016survey}.

\begin{figure}[tb]
  \centering
  \subfigcapskip = 0pt
  \subfigure[Regular intermittent computing.]{
    \label{fig:intermittent}
    \includegraphics[width=.99\linewidth]{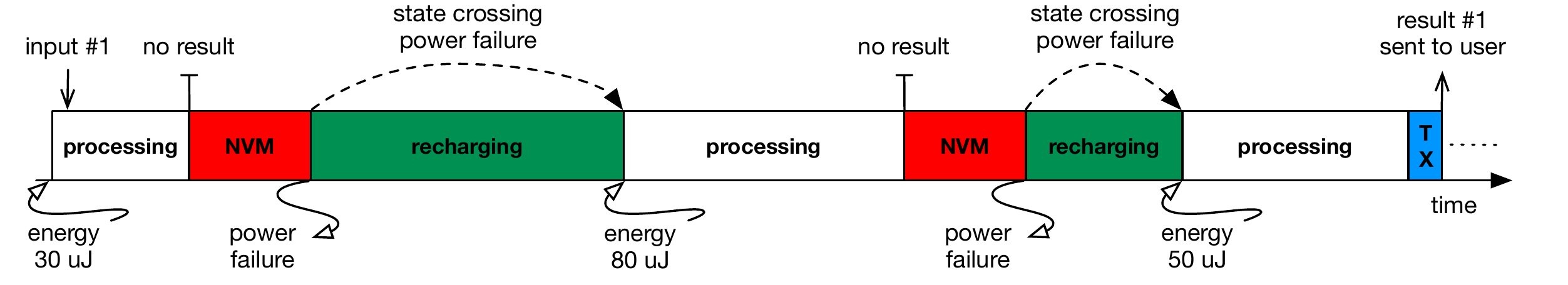}
  } 
  \subfigure[Approximate intermittent computing.]{
      \label{fig:approximate}
    \includegraphics[width=.99\linewidth]{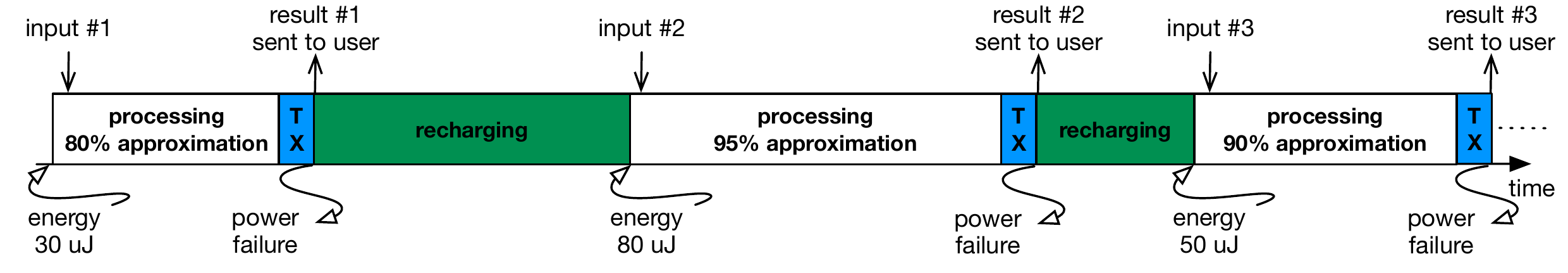}
  }
  \vspace{-3mm}
  \caption{Approximate intermittent computing trades the accuracy of the final result for the time and energy required to allow stateful computations to cross power failures. \capt{Such a shift allows applications to completely employ the available energy for useful application processing.}}
  \vspace{-3mm}
   \label{fig:intermittentVSapproximate}
 \end{figure}

These observations prompt us to establish a concept of \emph{approximate intermittent computing}.
In regular intermittent computing, shown in~\figref{fig:intermittent}, every input is processed until the precise outcome is eventually computed.
Because processing of an input rarely concludes before the first power failure, persistent state is employed to make stateful computations cross power failures.
This means that the latency to return a result to the user, for example, by transmitting a packet, extends over multiple power cycles and crucially includes the potentially large periods to recharge energy buffers.
While the system goes back and forth between main memory and NVM to dump and re-load application state for crossing power failures, many newer inputs are missed. 

In contrast, as shown in~\figref{fig:approximate}, approximate intermittent computing dictates that whenever a precise result cannot be obtained within a single power cycle, the system shall work towards an \emph{immediate approximate outcome}.
Approximation is the knob the system exploits to reduce the energy costs so that a result can be returned to the user \emph{before} the first power failure, \emph{shriking the extent of stateful computations to the single power cycle}.
Based on available energy, the system dynamically selects a level of approximation to process the current sample that ensures precisely this.
This means that, instead of seeking the most efficient way to make stateful computations cross power failures, as most existing solutions do, we effectively \emph{remove the problem} in the first place, as stateful computations are bound to conclude before the first power failure.

Our approach is fundamentally different, however, than the regular application of approximate computing techniques.
Indeed, application requirements normally dictate rigid lower bounds on the accuracy of results~\cite{mittal2016survey,han2013approximate}.
Approximate algorithms aim at ensuring this minimum accuracy by saving the greatest amount of resources, for example, energy.
In our case, the problem is effectively reversed.
Applications may not impose minimum accuracy requirements and be satisfied with whatever  is attainable, whereas \emph{a strict upper bound on energy consumption} exists due to the finite energy buffers.
Approximate algorithms must accordingly be designed to attain the greatest accuracy within a finite energy budget.

The ability to shrink the extent of stateful computations to the single power cycle entails that \emph{no persistent state needs to be carried over to the next power cycle}, and available energy is \emph{exclusively employed for useful computations}, rather than for NVM operations.
This means that:
\begin{glossy_enumerate}
\item the system is \emph{ready to process new inputs as soon as it restarts} after a power failure, as previous iterations necessarily concluded;
\item \emph{intermittence anomalies are not an issue}, because re-executions that combine volatile and non-volatile data do not occur;
\item \emph{handling peripheral states across power failures is unnecessary}, as computations always resume from the same point in the code;
  \item as there is no persistent state to maintain, devices \emph{need not be equipped with NVM} to run intermittent programs.
\end{glossy_enumerate}

The key trade-off we explore is the reduction in accuracy against the energy savings obtained by sparing the need of persistent state and the improvements in the latency to return the result to the user.
Latency is here, nonetheless, inversely proportional to system throughput.
The very notion of accuracy, however, along with the way it is measured and the definition of acceptable bounds, is inherently application-specific.
This is \emph{not only germane to our work}, but a general characteristic of approximate computing~\cite{mittal2016survey}.

\fakepar{Concrete cases} We first consider the case of human activity recognition~\cite{anguita2012human} using on-body acceleration and angular velocity sensors.
This application shows the characteristics discussed earlier and is an ideal candidate for energy harvesting~\cite{haroun2016investigation,romero2009kinetic}, as batteries are detrimental to user experience and increase a device's footprint.
Machine learning techniques, such as support vector machines~\cite{boser1992training}, are often employed to perform activity recognition~\cite{anguita2012human}.

We develop an \emph{anytime} variation of support vector machines, described in \secref{sec:asvm}, where accuracy of the classification is improved incrementally by processing one feature of the input samples at a time.
This is the knob that determines the level of approximation versus the energy cost of processing; the more features are processed, the more accurate is the classification, but also the higher is the energy cost.
Limiting the extent of stateful computations using anytime support vector machines requires to tie the number of processed features to the expected accuracy; we study this aspect analytically depending on the statistical nature of input data and the number of output classes.

Running classification tasks on resource-constrained embedded devices also requires additional efforts to fit a complex processing pipeline within a limited processing and memory budget~\cite{gobieski2019intelligence,islam2020zygarde}.
In \secref{sec:system}, we report on the prototype we build, including the customized hardware, the off-line data processing, and two alternative software implementations.
As we spare the need of persistent state, our prototype is the first device we are aware of to run real-world intermittent programs with no use of NVM.
The two implementations showcase the spectrum of possibilities offered by approximate intermittent computing.
One implementation employs available energy greedily, producing the most accurate result within the available energy budget.
The other implementation allows developers to set a lower bound in accuracy and postpones processing until the current energy budget ensures the required accuracy.

Our evaluation of approximate intermittent computing applied to human activity recognition,  reported in \secref{sec:eval}, is based on a combination of emulation and real-world experiments involving a total of 15 volunteers across a total of 24 days, producing roughly $\approx$842~hours of experiment data.
By comparing the performance in accuracy and system throughput, we show that approximate intermittent computing provides a $7x$ improvement in system throughput compared to state-of-the-art system support for regular intermittent computing, while retaining an 83\% accuracy in a setting where the best attainable accuracy is 88\%.
Moreover, with approximate intermittent computing, returning the result to the user always occur within the same power cycle \emph{by design}; with regular intermittent computing instead, the time when the classification is returned is entirely a function of energy patterns, extends across tens of power cycles, and includes the times to recharge the energy buffer.

Following the same path as in the case of human activity recognition, \secref{sec:discussion} demonstrates the applicability of approximate intermittent computing to a sharply different application, namely, embedded image processing.
We employ loop perforation techniques~\cite{mittal2016survey} as the knob to trade a loss of accuracy for a reduction in energy cost, allowing the system to return a result to the user before the first power failure, and again sparing the need of persistent state.
We build another hardware/software prototype, which we feed with diverse energy traces to explore the impact of various energy patterns on the accuracy of the output and energy consumption, compared to the same state-of-the-art system support for regular intermittent computing.
In this setting, approximate intermittent computing achieves a $5x$ improvement in system throughput, while providing an equivalent output in 84\% of the cases.

We release as open-source~\cite{opensource} the software artifacts and hardware schematics concurring to the results we present next, enabling others to continue exploring approximate intermittent computing.


\section{Related Work}
\label{sec:related}

Our combines intermittent computations with a novel use of approximate computing techniques.
We provide a brief account of existing efforts in these areas.

\fakepar{Intermittent computing} Most existing works in intermittent computing~\cite{batteryless-future} focus on how to make programs mantain the same semantics as a continuous execution, which requries stateful computations to cross power failures.
Common to these works is the use of some form of persistent state on NVM.
Two flavors exist.

Some solutions employ a form checkpointing ~\cite{mementos,Hibernus,Hibernus++,chinchilla,HarvOS,DICE,ratchet}.
This consists in replicating the application state on NVM at specific points in the code, where it is retrieved back once the system resumes with sufficient energy.
Systems such as Hibernus~\cite{Hibernus,Hibernus++} operate based on interrupts fired from a hardware device that prompts the application to take a checkpoint, for example, whenever the energy level falls below a threshold.
Differently, systems exist that place function calls in application code to proactively checkpoint~\cite{mementos,HarvOS,ratchet,chinchilla}.
The specific placement is a function of program structure and energy provisioning patterns.

Other approaches offer abstractions that programmers use to define and manage persistent state~\cite{dino,alpaca,chain,coala,Ink} and time profiles~\cite{mayfly}.
These approaches particularly target mixed-volatile platforms, while taking care of intermittence anomalies due to repeated executions of non-idempotent code~\cite{maioli21ewsn,brokenTM}.
For example, Alpaca~\cite{alpaca} defines tasks as individual execution units that run with transactional semantics against power failures and subsequent reboots, and channels to exchange data across tasks.

Approximate intermittent computing represents a different design standpoint.
Rather than retaining equivalence to continuous executions at the cost of making stateful computations cross power failures, we trade accuracy in the final results for better energy efficiency, making it possible to shrink the extent of stateful computations to the single power cycle.
Unlike a few existing works that explored similar directions~\cite{ganesan2019s,ma2017incidental}, we push this design to the point of taking away the need of persistent state, thus allowing systems to burn energy exclusively for useful application processing, while not requiring custom hardware.
Approximate intermittent computing is applicable where inaccurate results are tolerable, but is unfit for applications requiring precise computations, as discussed in \secref{sec:discussion}.

Additional issues arising in the execution of intermittent programs require testing the executions~\cite{maioli19lctes,cleancut,edb} and profiling their energy consumption~\cite{epic,SIREN}.
These techniques and tools are orthogonal to our work and may be applied to approximate intermittent programs as well.
We do rely on an existing energy estimation tool for intermittent programs in our prototypes of \secref{sec:system} and \secref{sec:discussion}.

\fakepar{Approximate computing} The need to reduce resource consumption in data-intensive applications originally motivates the development of approximate algorithms, namely, algorithms sacrificing the accuracy in the final result to gain in key performance metrics, such as processing times, memory occupation, or energy consumption.
A vast body of work exists on the subject~\cite{mittal2016survey,han2013approximate}.

From an algorithmic standpoint, our work is not different than most existing literature in approximate computing.
As the very notion of accuracy, the way it is measured, and the definition of acceptable bounds are application-specific, the specific data processing techniques we employ are also necessarily tied to a specific class of applications.
This is, in fact, one of the major limitations of approximate computing in general~\cite{han2013approximate}.
Nonetheless, these techniques are merely the specific instantiation of approximate intermittent computing we study here for the applications we consider.
Other instantiations are also possible that target different classes of applications, as we argue in \secref{sec:discussion}.

What is fundamentally different, however, is the formulation of the approximate computing problem, as we hinted earlier.
Instead of dealing with rigid lower bounds on the accuracy of results dictated by application requirements~\cite{mittal2016survey,han2013approximate}, we are to work with a fixed energy enveolope, dermined by the size of energy buffers.
In our case, algorithms must improve the accuracy obtained within the finite energy budget, whatever that may be, rather than improving resource usage, such as energy, within the constraint of a minimum accuracy bound. 
The traditional design of approximate algorithms is therefore often inapplicable~\cite{mittal2016survey,han2013approximate}, as the effort spent in computing is upper bound unlike in mainstream scenarios.

\fakepar{Output degradation in intermittent computing} Clostest to our work are systems that adapt the application execution based on available energy, possibly degrading the quality of the output tin situations of enerfgy scarcity.
One example is CatNap~\cite{catnap}, a programming model and run-time system that allows programmers to identify a subset of the code as time-critical.
When available energy is lower than expected, CatNap defers the execution of the non time-critical code to ensure timely execution of the time-critical one. If the schedule becomes unfeasible in situations of extreme energy scarcity, CatNap further degrades the quality of the output by either running different programmer-provided code for the same functionality, or the same code but less frequently. 

Works also exist that use multi-resolution inference and multi-exit strategies to handle a graceful degradation of the machine learning performance to meet temporal deadlines or depending on available energy.
For example, Sonic~\cite{gobieski2019intelligence} provides specialized support for intermittent executions, Zygarde~\cite{islam2020zygarde} aims at taking power failures into account to reduce the inference latency, whereas ePerceptive~\cite{eperceptive} builds upon Sonic by adding a dynamic multi-exit strategy. 
These works do represent initial steps in a similar direction as approximate intermittent computing, yet they do not seek to constrain the stateful processing steps within the same power cycle, and therefore still require the use of persistent state.


\section{Case in Point}
\label{sec:asvm}

We make a first concrete case for approximate intermittent computing using human activity recognition as a target application.
Our choice is not casual.
While human activity recognition is representative of the two characteristics discussed in the Introduction, it is also a nice fit for energy harvesting.
Kinetic energy abounds on human bodies~\cite{haroun2016investigation,romero2009kinetic}, whereas batteries are detrimental to user experience, as they increase a device’s footprint.

We specifically consider the framework of Anguita et al.~\cite{anguita2012human} for human activity recognition.
The application consists in classifying 50~\hz samples of acceleration and angular velocity readings from on-body sensors as representative of six possible human activities, including walking, walking upstairs, walking downstairs, standing, sitting, and laying.
Anguita et al.~\cite{anguita2012human} employ sensors on a smartphone and a regular support vector machine for classification, obtaining an absolute accuracy of 93.9\%.

Albeit the design of anytime support machines we describe next is originally motivated by human activity recognition, their applicability extends beyond that.
The underlying reasoning and mathematical properties \emph{are not}, indeed, tied to the specific application.

\begin{figure}[tb]
  \centering
  \includegraphics[width=.6\linewidth, angle=-90]{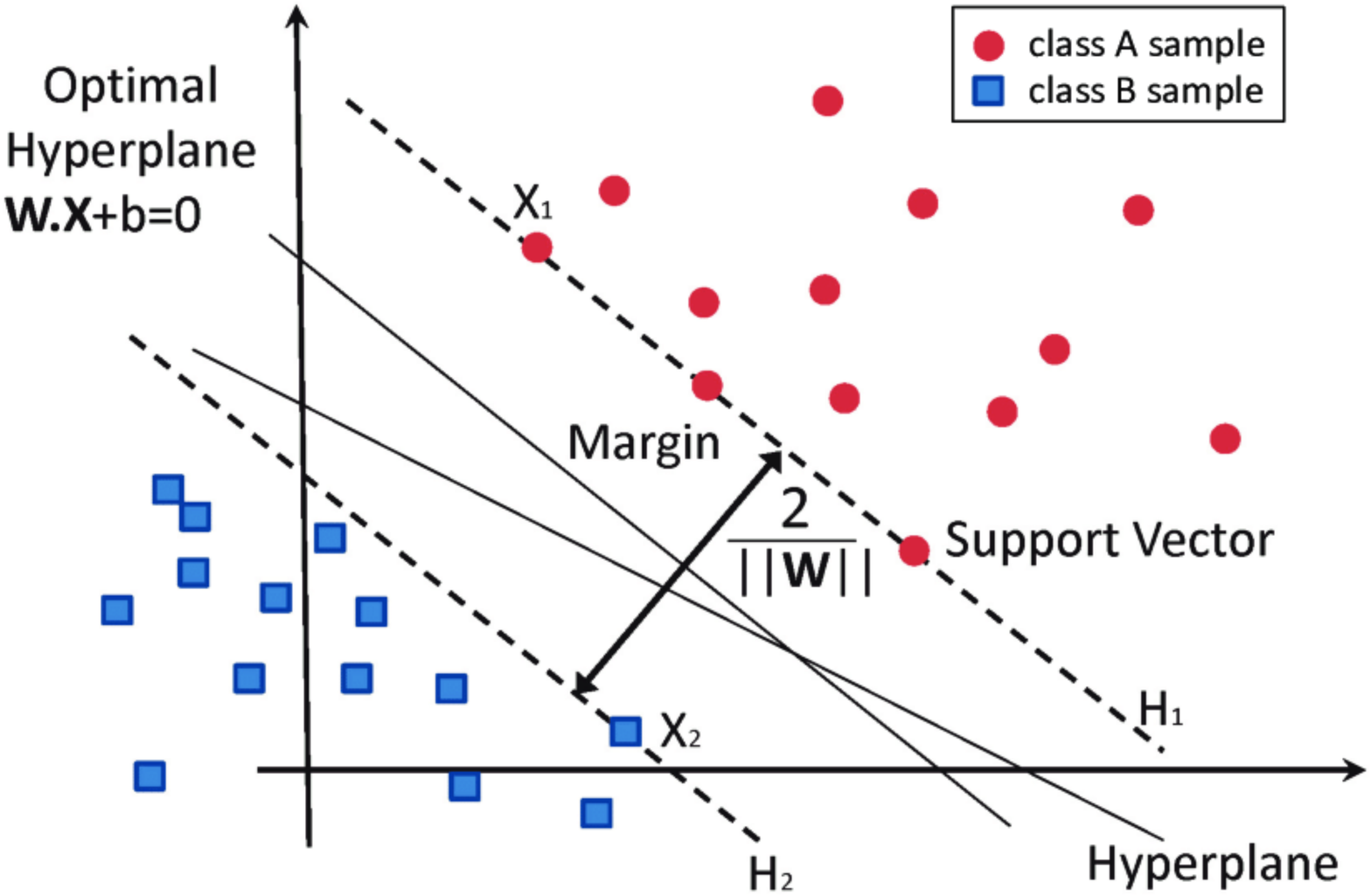}
  \vspace{-2mm}
   \caption{Geometrical representation of separating hyperplane and support vectors in two dimensions.}
  \vspace{-2mm}
   \label{fig:svm}
\end{figure}

\subsection{Preliminaries}

We provide necessary background to later understand our design.

\fakepar{Support vector machines} A geometrical analogy helps understand support vector machines, shown in \figref{fig:svm}.
Consider the case of only two possible classes $A$ and $B$, for simplicity.
Support vector machines use training data to identify a hyperplane that provides the best distinction between the two classes.
The hyperplane is the one that is maximally distant from the closest vector representing the input sample.
These vectors are called support vectors.
The separating hyperplane is commonly identified by solving a quadratic programming problem~\cite{mohri2018foundations} using training data.
The computational complexity of training is $O(n^3)$, with $n$ being the cardinality of the training set.
The output of the training phase is a tuple of coefficients with the same dimensionality as the input samples.

Support vector machines rely on the assumption that the data set is linearly separable, that is, at least one hyperplane $W $ exists such that all the input samples in $A$ are on one side of $W$, and all input samples in $B$ are on the other.
This is not necessarily the case.
If the data set is not linearly separable, the data is projected in high-dimensionality spaces where it becomes so, using kernel functions~\cite{mohri2018foundations}.
These retain the applicability of support vector machines in cases where the original data is not linearly separable, but incur in additional memory consumption and processing.

Support vector machines are usable also when the classification task extends to multiple classes.
Two methods are available.
One-versus-rest (OvR) methods classify input samples as belonging to the class whose hyperplane is the farthest away.
Instead, one-versus-one (OvO) methods use one classifier for each pair of classes $C_i,C_j$.
The class that matches most frequently among all possible pairs is returned as output.
As OvO methods require one hyperplane for each pair of classes and a corresponding comparison when classifying input samples, they incur in greater processing and memory cost in both training and classification.
OvR methods are therefore usually favored on embedded devices.

\fakepar{Approximation for support vector machines} Approximate variations of support vector machines exist. 
For example, Decoste et al.~\cite{decoste2002anytime} design interval-valued support vector machines.
Whether an input sample~\vect{x} belongs to class $C_1$ or $C_2$ may be computed by determining the sign of
\begin{equation}
  \label{eq:sign}
  \sum_{p \in C_1}d^2_{xp} - \sum_{q \in C_2}d^2_{xq},
\end{equation}
where $d^2_{xp}$ is the squared distance between the input sample \vect{x} and the support vector~\vect{p}.
Instead of computing \eqref{eq:sign} directly, Decoste et al.~\cite{decoste2002anytime} present an algorithm to compute corresponding bounds that are continuously narrowed until the sign of \eqref{eq:sign} is determined.

Wagstaff et al.~\cite{wagstaff2010progressive} train two different models, one that is extremely accurate but incurs in high processing cost during classification, the other that is less precise but extremely simple to use.
Input samples are first classified using the simple model, and the probability that the classification is correct is accordingly computed.
In cases this falls below a threshold, the accurate model is used for re-running the classification.

Both approaches are, however, largely inapplicable for approximate intermittent computing.
Interval-valued support vector machines incur, in the worst case, in greater processing cost than regular support vector machines~\cite{decoste2002anytime}.
As this is proportional to energy consumption on the devices we target, this characteristic defeats the very purpose of using approximations.
The approach of Wagstaff et al.~\cite{wagstaff2010progressive} offers little flexibility, as there are only two possibilities for tuning the classification accuracy: one is either happy with the simple model, or runs the accurate one as well.


\subsection{Anytime Support Vector Machines}
\label{sec:asvmspecs}

Like Anguita et al.~\cite{anguita2012human}, we use the OvR method.
Say \vect{w_1}, \vect{w_2} $\dots$ \vect{w_c} are the vectors representing the hyperplanes of the $c$ classes.
An input sample \vect{x} may be classified by identifying the class corresponding to the hyperplane that yields the largest inner product with \vect{x}.
Our claim is that we can achieve an approximate classification by using fewer features than the $n$ available, that is
\begin{equation}
  \label{eq:approx}
  \vect{w_ix} = \sum^n_{j=1}w_{ij}x_{j} \approx \sum^p_{j=1}w_{ij}x_{j}, 
\end{equation}
where $1<i<c$ and $p < n$.
Intuitively, we use only a subset of the available features to compute the classification.
The closer is $p$ to $n$, the more accurate is the classification, and viceversa.

The classification may then be computed incrementally, by ca\-ching approximate results and adding more features as energy is available.
This is particularly beneficial whenever the additional features are
not immediately available from sensor data, but require additional processing on the latter before they can be used for classification.
By limiting the classification step to a subset of the features, we
save the processing overhead of not just the classification step
itself, but also of the computation required to compute the additional
features form sensor data.

To employ this form of approximation in intermittent computing, the key question is how to tie the number of features we do process to the expected accuracy in the final classification.
This essentially represents the trade-off between energy cost and accuracy.
In the general case, we are interested in computing the probability that a classification using $p < n$ features is coherent with the one obtained using $n$ features, that is, it is the same as the most accurate classification we can achieve.
  We want to study this probability as a function of $p$.
  We call $\mathit{class}_{mi}$ the classification of the $i$-th sample using $m$ features.
  Therefore, we are to calculate, depending on $p$ 
\begin{equation}
P(\mathit{class}_{pi}=\mathit{class}_{ni}).
  \end{equation}

How to compute this probability depends on whether \emph{i)} the features are independent or correlated, and \emph{ii)} the classification is binary or extends to multiple classes.
We briefly illustrate next the simplest case to illustrate the underlying intuition; the analytical details and derivations for all other cases are also available~\cite{techrep}.

Let us consider two classes and independent features.
When using all $n$ available features, the classification of an input sample \vect{x_i} is determined by the sign of 
\begin{equation}
  \label{eq:classify2precise}
S_i = \vect{w}\vect{x_i} = \sum^n_{j=1}c_jx_{ij},
  \end{equation}
  where $\vect{w} = [c_{1}, c_{2}, \dots, c_{n}]$ is the vector representing the hyperplane used to classify the input samples, and $\vect{x_i} = [x_{i1}, x_{i2}, \dots, x_{in}]$ is the $i$-th input sample we process.

An approximate classification for the $i$-th input sample may be obtained by computing the sign of  
\begin{equation}
  \label{eq:classify2approx}
S_{ip} = \vect{w}\vect{x_i} = \sum^p_{j=1}c_jx_{ij}, \quad p < n.
\end{equation}
Note that the sign of \eqref{eq:classify2precise} is coherent with that of \eqref{eq:classify2approx} as long as
\begin{equation}
  \label{eq:classify2coherent}
S_{ip} \geq -R_{ip} = -\sum^n_{j=p+1}c_jx_{ij}, 
\end{equation}
where $R_{ip}$ represents the contribution to $S_i$ of the features we are \emph{not} considering.
Intuitively, \eqref{eq:classify2coherent} states that such contribution is not sufficient to flip the sign of $S_{ip}$ compared to $S_i$, and hence the two classifications are coherent.

\eqref{eq:classify2coherent} also suggests what are the features that should be processed first.
In fact, for the same input sample, $R_{ip}$ is as small as the features $p+1, \dots, n$ correspond to smaller coefficients~$c_j$.
This entails that the ideal order to process features looks at the magnitude of the corresponding coefficients in the separating hyperplane.
Features with larger coefficients bear a stronger contribution in determining the final classification, and are therefore those we should process first.
We confirm this observation experimentally in \secref{sec:eval}.

In the case of independent and normally distributed coefficients $c_{1}, c_{2}, \dots, c_{n}$, we eventually derive that
\begin{equation}
 \label{eq:classify2coherent-normal}  
  P(\mathit{class}_{pi}=\mathit{class}_{ni}) = 2 \int_{k=0}^{\infty}f_{S_{ip}}(k)(1-F_{R_{ip}}(k))dk,
  \end{equation}
  where $f_{S_{ip}}$ and $F_{R_{ip}}$ may be determined numerically~\cite{techrep}, making \eqref{eq:classify2coherent-normal} cheap to compute.

  The case of multiple classes follows as a natural extension. Say $C_1, C_2, \dots, C_c$ are the possible classes, and \vect{w_1}, \vect{w_2} $\dots$ \vect{w_c} are the vectors representing the corresponding hyperplanes.
  For a generic class $h$, \eqref{eq:classify2precise} may be extended to the case of multiple classes as 
  \begin{equation}
  \label{eq:classifyMprecise}
S_{hi} = \vect{w_h}\vect{x_i} = \sum^n_{j=1}c_{hj}x_{ij}.
\end{equation}

The input sample $i$ is classified as belonging to class $C_h$ when using all $n$ available features such that
  \begin{equation}
  \label{eq:classifyMargmax}
\mathit{class}_{ni} = \mathit{argmax}_h(S_{hi}), \quad 1 \leq h \leq c.
\end{equation}

It is possible to repeat the same reasoning of \eqref{eq:classify2coherent} for each individual class $C_h$ compared to all others.
Therefore, in the case of independent and normally distributed coefficients, the probability that the classification using $ p < n$ features is coherent with the one obtained using $n$ features is given by \eqref{eq:classify2coherent-normal} for a generic class $C_h$, multiplied by the probability that $h$ is precisely the one solving \eqref{eq:classifyMargmax}.
In this case as well, we eventually derive an expression that may computed numerically~\cite{techrep}.
It is also possible to derive similar expressions in the case of correlated coefficients, by taking into account the corresponding covariance matrix~\cite{techrep}.
In this case too, the value of the expressions may be computed numerically.



\section{Prototype}
\label{sec:system}

We describe next the hardware we build, the data processing to extract features useful for classification, how we train the models, and how we implement two alternative classification pipelines on resource-constrained devices. 

\subsection{Hardware}


\begin{figure}
    \includegraphics[height=3.5cm]{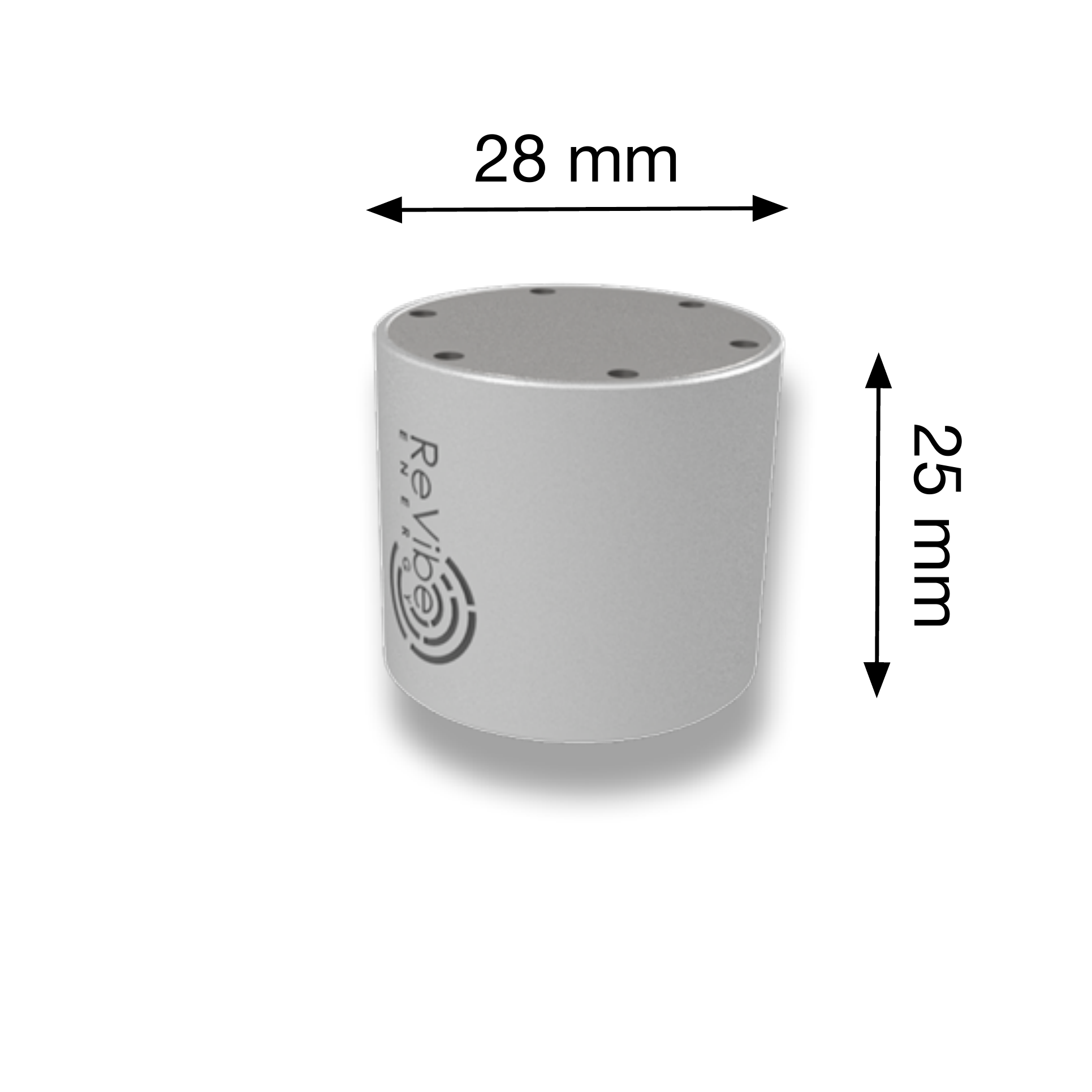}
  \vspace{-8mm}
   \caption{ReVibe modelQ kinetic energy harvester.}
   \label{fig:revibe}
\end{figure}

We manufacture a custom board hosting an MSP430-FR5659 MCU to perform human activity recognition using kinetic energy.
The MCU is equipped with 64\kb of volatile RAM together with 512\kb of FRAM as NVM.
We use the FRAM exclusively when running regular intermittent computing techniques. 
Acceleration and angular velocity readings are obtained through an Analog Devices ADXL362
accelerometer and an STM Electronics L3GD20H gyroscope respectively, both connected through SPI. 
The board features BLE connectivity through a Nordic nRF51822 chip.

For energy harvesting, we use a ReVibe modelQ~\cite{revibe-modelQ} kinetic transducer,  shown in \figref{fig:revibe}, which we order with a customized resonance frequency based on the spectral profile of raw accelerometer data we gather in a short pre-deployment trial.
We choose the modelQ over alternatives, for example, the modelD~\cite{revibe-modelD} of the same manufacturer, because of the smaller form factor and the higher power output at the target frequencies.

Similar to existing deployments of intermittent computing~\cite{flicker,permadaq,sensys20deployment}, the harvester is attached to a BQ25505 combined booster and buck converter that charges a 1470\uf capacitor we use as energy buffer.
We determine its size through a mixed analytical and experimental approach~\cite{teg-apps}, striking a balance between charging times and available energy.
A too large capacitor may take long to charge to a sufficient level, yielding large periods of no system operation.
A too small capacitor may not suffice to supply enough energy for worst-case processing scenarios.
Whenever required, the capacitor's current charge is read through an Analog Devices ultralow-power LTC1417 analog-to-digitial converter.

We create 12 identical hardware prototypes.
The device is admittedly large, but still wearable with no major issues by the volunteers involved in the trial of \secref{sec:eval}.
Energy-harvesting wearable devices exist in much smaller form factors~\cite{teg-apps,magno2016infinitime}.
Our prototype is mainly meant as a platform for evaluation, not as an end-user product.

\subsection{Data and Training}

We sample sensors at 50\hz until we accumulate a window of~.2\secs of acceleration and angular velocity readings.
These data as a whole represent the individual sensor sample processed for classification. 
We use a 3rd order Butterworth filter with a cutoff frequency of 20\hz to remove the noise, as 99\% of the signal energy is found below that~\cite{anguita2012human}.
A second low-pass filter accounts for gravity.

Out of the raw data, Anguita et al.~\cite{anguita2012human} compute a total of 561 features for training and classification.
Not all of these features are, however, generating linearly separable samples, therefore demanding the use of kernel functions.
Because of the increased overhead due to these, we limit ourselves to the 140 signal features that generate linearly separable samples.
Together with the specific implementation techniques we employ, this also ensures that in a continuous execution, all 140 features may be used for classification before the new sensor readings are gathered.
This represents the most accurate classification we can possibly provide.
The features we compute range from simple window operators such as average and standard deviation, to sophisticated ones such as fast Fourier transforms and spectral density distributions.

We use the original dataset of Anguita et al.~\cite{anguita2013public} for training.
Learning is accomplished in the regular way, using the SVM Python library from the \code{scipy} package.
Based on \secref{sec:asvmspecs}, with the same training set we numerically compute the probability that a classification using only $p<140$ features is coherent with the classification obtained with all 140 available features.
For each feature, we use energy estimation tools for intermittent computing~\cite{epic} to profile the energy necessary to add that specific feature to the existing classification.
Such an energy cost is fixed for a feature, but varies across features mainly because of the processing to extract the feature from the raw sensor readings.

The entire data processing, training phase, accuracy estimation depending on the $p$ features, and energy profiling run on a standard desktop machine in a fully automated way.
Processing the data set of Anguita et al.~\cite{anguita2013public} requires less than an hour.

\subsection{Software}

We implement the classification pipeline using C/C++.
The classification process starts as soon as a new window of sensor samples is gathered.
We create two implementations:

\begin{description}
\item[\textsc{Greedy}.] The \greedy implementation continues to add features to the existing classification, progressively refining the accuracy, until either just the right amount of energy is left to send out a BLE packet with the 1-byte output, or all available features are used.
  In the latter case, a BLE packet is generated thereafter and the node switches to the lowest-power mode available that allows the system to wake up again in one minute for the next iteration of sensor sampling.
  No issue arises if the system dies because of energy depletion at this stage; the result is already returned to the user.
\item[\textsc{Smart}.] Based on the information provided by the offline phases, the \smart implementation first determines whether the available energy is sufficient to achieve a classification accuracy above a user-defined threshold $A$ and finds in a look-up table the corresponding number $p'$ of features to be used.
  If energy is insufficient, it skips this round of classification and switches to the lowest-power mode, similar to \greedy, waiting for the next sensor samples.
  Otherwise, it immediately uses all $p'$ samples and then switches to \greedy mode.  
\end{description}

Note that the operation of the \smart implementation ensures that the user-defined threshold is met for all input samples that are actually processed.
At the same time, it also ensures that any energy left or obtained while running is employed to further refine the accuracy of the classification before returning it to the user.

Both implementations employ fixed-point arithmetics due to the lack of hardware support for floating-point operations on the MCU we target and space-efficient data structures to store \emph{i)} the learned models output by the training phase, \emph{ii)} the mapping between the $p$  processed features to the expected classification accuracy, and \emph{iii)} information on the energy cost to add the $i$-th feature to the existing classification.
These information occupy $\approx$18\kb of the $64$\kb of available memory, leaving ample room for additional functionality if necessary.










\section{Evaluation}
\label{sec:eval}

We consider four key metrics.
The \emph{accuracy} of classification represents the matching between recognized human activities and ground truth, whenever available.
This metric tops at the accuracy provided by a continous execution running without interruptions, which always uses all available features.
Whenever ground truth is not available, we measure the \emph{coherence} of the classification returned by approximate intermittent computing against continous executions or regular intermittent computing, as discussed in \secref{sec:asvm}.
The system \emph{throughput} measures the number of classifications returned to the user throughout an experiment duration, whereas the \emph{latency} indicates when those classifications are emitted compared to when the sensor samples are acquired.

For approximate intermittent computing, we study the performance of the \smart implementation, using a 80\% or 60\% lower bound in accuracy, and of \greedy, as described in \secref{sec:system}.
In addition to a continous execution, we use Chinchilla as a baseline~\cite{chinchilla}.
Chinchilla overprovisions code with checkpoints to ensure forwrd progress with scarce energy, then dynamically disables checkpoints to adapt to situations of energy abundance.
Because of this, Chinchilla efficiently matches the varying energy levels in our target scenarios.
We also clock the MCU at 8\mhz  to avoid wait states when writing or reading checkpoints on FRAM.  
The performance we measure for this baseline thus represents a best case.

We study the relevant trade-offs from multiple angles, using different tools and settings:
\begin{enumerate*}
\item in \secref{sec:emulationAccuracy}, we use emulation experiments to compare the expected and measured accuracy of classification as a function of the number of features used for classification, providing quantitative support to the analysis in \secref{sec:asvmspecs};
\item in \secref{sec:emulationAll}, we use emulation experiments to compare the latency, accuracy, and throughput of the implementations in \secref{sec:system} against either a continuous execution or Chinchilla; 
\item in \secref{sec:realContinous}, we involve six volunteers for about 56 hours each to compare the latency, accuracy, and throughput of the implementations in \secref{sec:system} against  a continuous execution, using two identical devices on the same person's wrist;
\item in \secref{sec:realIntermittent}, we use the same setup as the previous case with another six volunteers for about 58 hours each to run a comparison against an implementation using Chinchilla.
\end{enumerate*}

The volunteers we involve include senior members of our lab and their spouses or parents.
The diversity of activities they are involved in, ranging from coding or studying to driving or exercising, caters for a range of different settings.
The experiments in \secref{sec:emulationAccuracy} and \secref{sec:emulationAll} are enabled by labeled sensor data and energy traces we collect with three of these volunteers using a battery-powered version of the prototype of \secref{sec:system} for about 56 hours each.
Note that these data is different from the dataset used for training.
The emulation experiments use an extension of the MSPSim emulator~\cite{SIREN,eriksson2009cooja} that provides support for using FRAM as NVM and accounts for the corresponding energy consumption.
Experiments using the real prototype output the classification over BLE to a smartphone the user carries.
The continuous executions are obtained with the same battery-powered version of our prototype mentioned earlier.

  

\subsection{Expected Accuracy}
  \label{sec:emulationAccuracy}

  \begin{figure}[tb]
  \centering
  \includegraphics[width=\linewidth]{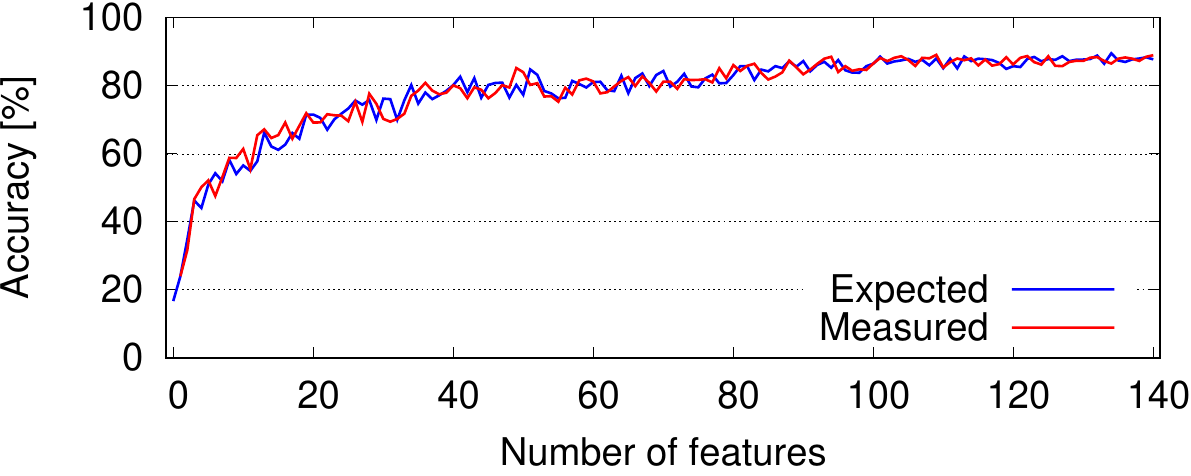}
  \vspace{-4mm}
  \caption{Emulation experiments: expected and measured accuracy as a function of the number of features used for classification. \capt{The expected accuracy computed according to \secref{sec:asvmspecs} is constantly close to the measured accuracy. The analysis of \secref{sec:asvmspecs} can, indeed, be used to forecast the accuracy as a function of the number of features used for classification. Using all available features, the accuracy is around 88\%, in line with the results of Anguita et al.~\cite{anguita2012human}.}}
  \label{fig:expectedAccuracy}
  \vspace{-3mm}
\end{figure}

The analysis of \secref{sec:asvmspecs} provides a conceptual and quantitative basis for the application of approximate intermittent computing.
Here we check that this analysis can indeed be relied upon, using the labeled data we collect, including ground truth.

\figref{fig:expectedAccuracy} shows the results of our emulation experiments comparing the expected and measured accuracy of classification, as a function of the number of processed features.
The expected accuracy, computed based on the analysis of \secref{sec:asvmspecs}, is constantly close to the measured accuracy.
The delta between the curves also appears largely independent of the number of features processed, providing evidence of the general applicability of our analysis.

  \begin{figure}[tb]
    \centering
    \vspace{-2mm}
  \subfigure[{Classification accuracy compared to ground truth [\%].}]{
    \label{fig:emulationAccuracy}
   \includegraphics[angle=270,width=.8\linewidth]{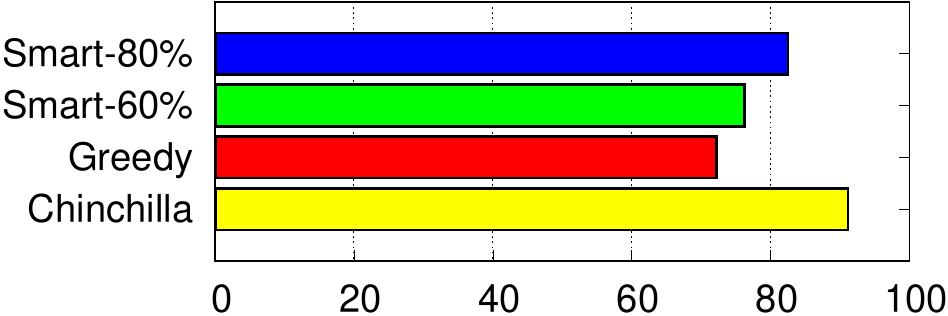}
  }
  \subfigure[{System throughput normalized to continuous execution [\%].}]{
    \label{fig:emulationThroughput}
   \includegraphics[angle=270,width=.8\linewidth]{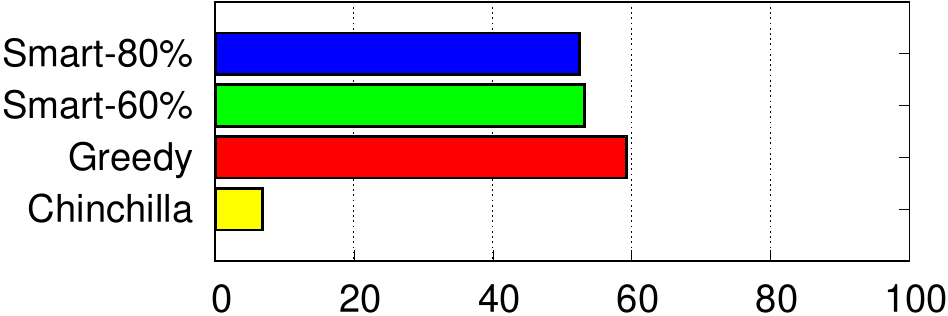}
 }
   \vspace{-4mm}
   \caption{Emulation experiments: classification accuracy and system throughput normalized to a continuous execution. \capt{The \greedy implementation captures energy fluctuations most efficiently, improving system throughput at the expenses of slightly lower accuracy. The \smart implementations ensure a lower bound in accuracy for every processed sample, at the cost of a slight reduction in throughput. Using Chinchilla provides the best possible classification accuracy, but must invest significant energy in handling persistent state, severely impacting the throughput.}}
  \label{fig:emulationThroughputAccuracy}
\end{figure}

\begin{figure*}[tb]
  \centering
  \includegraphics[width=\linewidth]{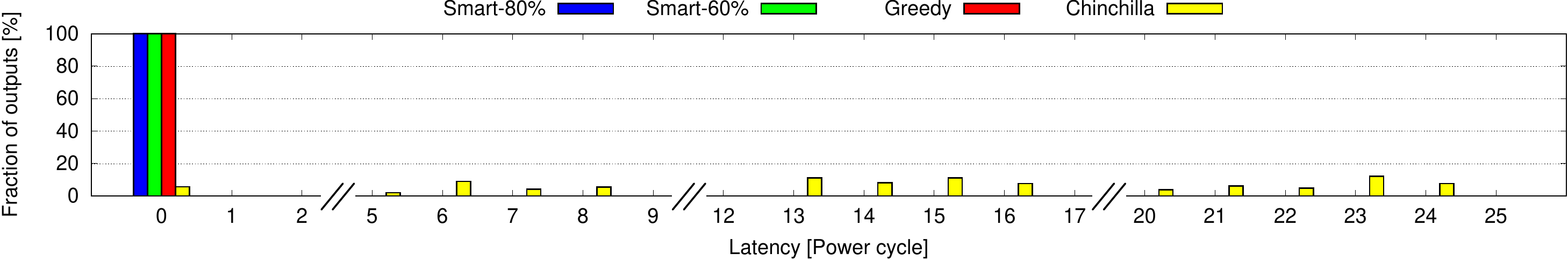}
  \vspace{-5mm}
  \caption{Emulation experiments: distribution of the latency to return the classification, measured in number of power cycles between when sensor data is acquired until the classification is emitted. \capt{Approximate intermittent computing always returns the result to the user within the same power cycle. Because intermittent computing uses persistent state to cross periods of energy unavailability until all available features are processed, the time when the classification is returned is entirely a function of energy patterns.}}
  \vspace{-2mm}
  \label{fig:latencyEmulation}
\end{figure*}

\figref{fig:expectedAccuracy} also offers a more general opportunity to gain a qualitative insight into the behavior of approximate intermittent computing.
As expected, the blue curve starts at 16,6\% because with no features, determining the correct classification equates to a random event with uniform distribution over the six possible classes.
As the number of features we process increases, both curves rapidly grow.
The first few features, which in our case come from processing the FFT of the input signal, significantly contribute to improving the accuracy of classification.
The curves eventually flatten out as the contribution of the latest features we process only marginally improves the obtained accuracy.
Both expected and measured accuracy top at around 88\%.
This is in line with the results of Anguita et al.~\cite{anguita2012human}, who obtain a 93.9\% accuracy using many more features compared to our system and the same training data.

\subsection{Comparing with Ground Truth}
\label{sec:emulationAll}

Using labeled data, we replay the execution of either implementation of approximate intermittent computing, as well as of the continuous baseline and of the one using Chinchilla, using the exact same sensor data and energy traces.

\figref{fig:emulationThroughputAccuracy} illustrates the results in accuracy and throughput.
\figref{fig:emulationAccuracy} shows how Chinchilla provides, for the single input sample, the best possible accuracy as it always uses all available features.
\figref{fig:emulationThroughput} illustrates, however, that doing so comes at a tremendous cost in terms of throughput, because Chinchilla stretches the processing of every single samples across multiple power cycles, missing the opportunity to process new samples.
Approximate intermittent computing, in contrast, trades a limited loss of accuracy for much greater throughput.
One of the fundamental benefits is apparent here: the energy budget is entirely spent for useful application processing, rather than for managing persistent state on the energy-hungry NVM.  
The loss of accuracy is around 11\% worst-case, yet the throughput improvements reach up to $7x$ that of Chinchilla.

Comparing either configuration of \smart with \greedy, \figref{fig:emulationAccuracy} shows a slightly higher accuracy for the former.
This is due to those samples that, in situations of energy scarcity, \smart discards as the number of features the system can process is too limited to match the required lower bound.
\greedy proceeds anyways by computing the classification with fewer features than usual, likely obtaining a less accurate output.
The effect of this is also apparent in \figref{fig:emulationThroughput}.
The samples that \smart decides to drop do not produce an output, causing a reduction of the throughput.
The same observations apply also between the two configurations of \smart, as a higher lower bound for accuracy improves the latter at the expense of additional dropped samples and therefore lower throughput.

Note how \figref{fig:emulationThroughput} also generally shows the impact of using ambient energy, in that energy harvesting causes a device not to run as often as a battery-powered one that can afford to execute continuously.
This is essentially the price to pay for a battery-less system.
Still, using approximate intermittent computing, more than half of the classifications that a continous execution would produce are indeed returned by the user when using kinetic energy.
Let apart experiences using solar radiation as the energy source, these results generally represent a significant improvement compared to existing deployments of intermittent computing~\cite{sensys20deployment,flicker,teg-apps}.

The improvements in throughput stem from the reduced latency to return the result to the user in approximate intermittent computing.
\figref{fig:latencyEmulation} illustrates this metric, measured in the number of power cycles, that is, the number of times the device wakes up with new energy, since the sensor data is acquired and until the classification is emitted.
The implementations using approximate intermittent computing return the classification within the same power cycle.
This is obtained by design, as the number of features we process is tuned for returning the output before the first power failure.
In contrast, Chinchilla is at the mercy of the energy source; computations stop and resume, using persistent state, until the ambient cumulatively provides sufficient energy to process all available features.
The latency therefore covers also the periods for recharging the energy buffer and no clear pattern emerges.
A non-negligible fraction of the outputs are even returned tens of power cycles later than when the sensor data is acquired.

\subsection{Comparing with Continous Executions}
\label{sec:realContinous}

We use two identical prototypes on the same person's wrist.
One of them is running either of the approximate intermittent computing implementations of \secref{sec:system}, the other one is battery-powered and executes continuously.
With six volunteers, we run two instances of every approximate intermittent computing implementation and six continous executions we compare with.
We align the power cycle information with the continous execution based on the interval the BLE packets are received at the user smartphone.
Should this interval be lower than .2\secs, which equals the duration of sensor sampling, we consider the two classifications to be aligned.

  \begin{figure}[tb]
  \centering
  \subfigure[{Classification coherence compared to continous execution  [\%].}]{
    \label{fig:coherenceContinous}
   \includegraphics[angle=270,width=.8\linewidth]{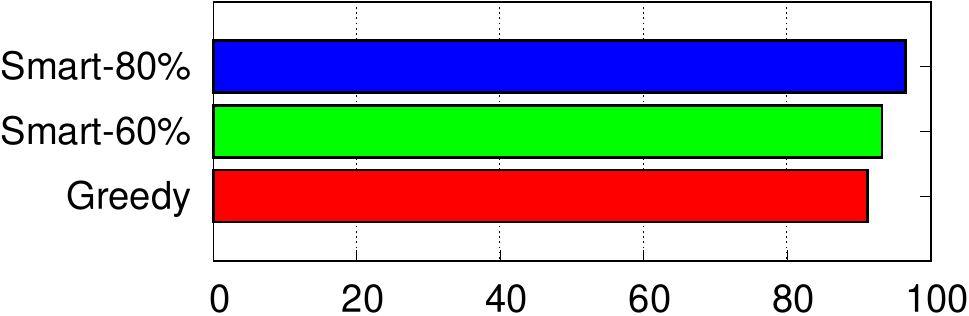}
  }
  \subfigure[{System throughput normalized to continuous execution [\%].}]{
   \label{fig:continousThroughput}
   \includegraphics[angle=270,width=.8\linewidth]{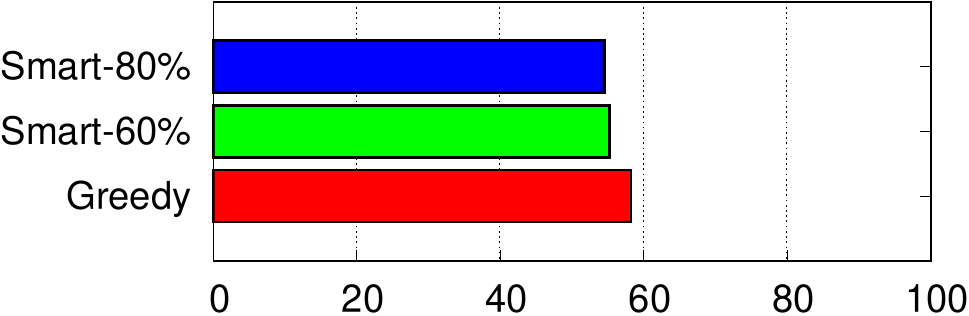}
 }
   \vspace{-3mm}
   \caption{Real-world experiments: coherence of the classification of approximate intermittent computing against continous executions and system throughput normalized to a continuous execution. \capt{Approximate intermittent computing returns the same classification of a continous execution in the majority of the cases. More than half of the samples processed by a continuous execution are processed by approximate intermittent computing too.}}
     \vspace{-3mm}
  \label{fig:continousThroughputCoherence}
\end{figure}

\figref{fig:coherenceContinous} shows the results we obtain in coherence of the classification.
This time we cannot reason on absolute accuracy as in \secref{sec:emulationAll}, because ground truth is not available.
In at least 91.2\% of the cases, however, the classification of human activities returned by approximate intermittent computing is the same as in a continous execution.
However, approximate intermittent computing runs in a completely self-sustained manner, using kinetic energy.
 The coherence is higher for \smart because of the reasons explained earlier: the lower bound on expected accuracy makes \smart discard samples in situations where the (too) little available energy would yield a less accurate classification.

 Unsurprisingly, \figref{fig:continousThroughput} shows again that relying on ambient energy severely impacts system throughput,
These results are in line with the emulation results of \figref{fig:emulationThroughput}, giving us confidence on the correctness of the setup.
Compared to \figref{fig:coherenceContinous}, the relative trends between the different implementations are reversed; \greedy shows higher throughput as it opportunistically consumes energy whenever available, returning more results at the expense of lower accuracy.
We do not report on latency here, as all implementations we test here return the classification within the same power cycle.

\subsection{Comparing with Chinchilla}
\label{sec:realIntermittent}

We use the same setup as in \secref{sec:realContinous}, but replace the continous executions with an implementation that uses Chinchilla.
Because the two devices on a person's wrist are exposed to the same movements, we verify that their energy patterns are almost identical.
To align power cycle information with Chinchilla, it is sufficient to check what checkpoints are taken since the sensors samples are acquired and until the result is transmitted.
This allows us to compute how many power cycles in the past the samples originate from.
  
  \begin{figure}[tb]
  \centering
  \subfigure[{Classification coherence compared to Chinchilla [\%].}]{
    \label{fig:coherenceIntermittent}
   \includegraphics[angle=270,width=.8\linewidth]{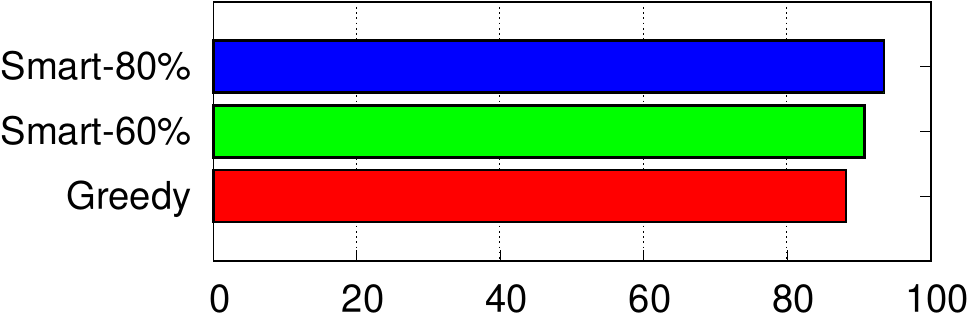}
  }
  \subfigure[{System throughput normalized to \greedy [\%].}]{
    \label{fig:intermittentThroughput}
   \includegraphics[angle=270,width=.8\linewidth]{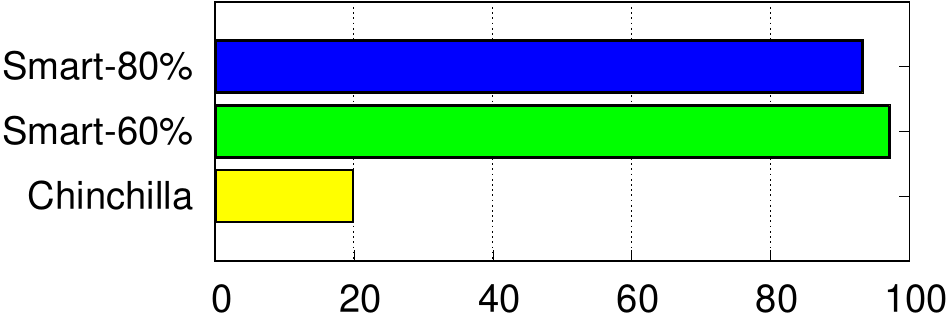}
  }  \vspace{-3mm}
  \caption{Real-world experiments: coherence of the classification of approximate intermittent computing against Chinchilla, and system throughput normalized to that of \greedy. \capt{The performance in coherence mirrors \figref{fig:coherenceContinous}, as Chinchilla processes all available samples like a continous execution. Throughput enabled by Chinchilla is much lower than approximate intermittent computing, as the former misses many opportunities to process newer samples.}}
    \vspace{-2mm}
\label{fig:intermittentThroughputCoherence}
\end{figure}

\figref{fig:coherenceIntermittent} shows the coherence of the classification obtained by the implementations of approximate intermittent computing compared to Chinchilla.
The results here mirror those of \figref{fig:coherenceContinous}.
The implementation using Chinchilla exploits all available features anyways, exactly like the continous execution.
Therefore, the accuracy it achieves is the same as a continous execution and thus the coherence with the classification of approximate intermittent computing is also similar.
What is different, however, is the number of sensor samples that Chinchilla manages to process.
As this implementation must cross periods of energy unavailability, using persistent state, processing of a single sample extends across multiple power cycles, preventing the the acquisition of newer samples.

\begin{figure*}[tb]
  \centering
  \includegraphics[width=\linewidth]{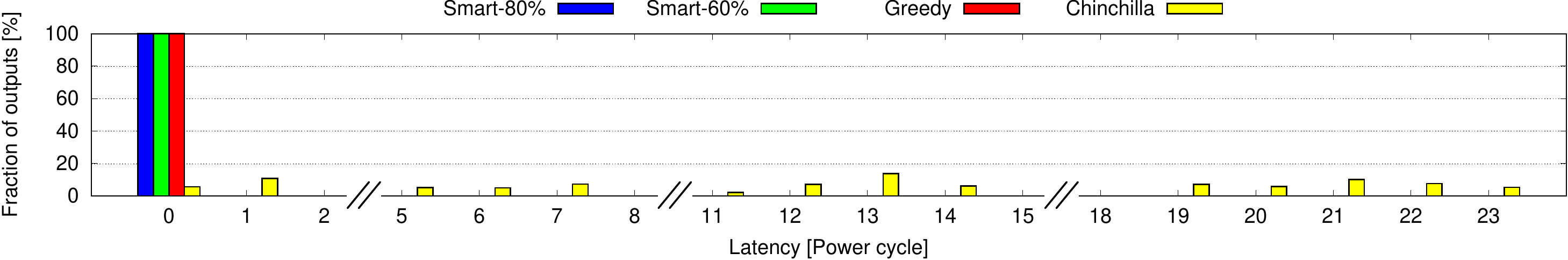}
    \vspace{-3mm}
    \caption{Real-world experiments: distribution of the latency to return the classification, measured in number of power cycles between when sensor data is acquired until the classification is emitted. \capt{Unlike approximate intermittent computing that returns the classification before the first power failure, Chinchilla stretches the processing across multiple power cycles.}}
    \vspace{-2mm} 
  \label{fig:latencyIntermittent}
\end{figure*}

One consequence of this is a reduction in system throughput, as shown in \figref{fig:intermittentThroughput}.
This time the data is normalized to the best performing implementation among the ones we test, which is \greedy.
Chinchilla can only provide a fraction of the results that approximate intermittent computing can dispense.
As seen before, improved throughput is enabled by reduced latency to return the result to the user, shown in \figref{fig:latencyIntermittent}.
By design, approximate intermittent computing returns the classification before the first power failure.
The time it takes for Chinchilla to return the result is a function of energy patterns, similar to \figref{fig:latencyEmulation}.
This latency stretches across even tens of power cycles and includes the periods for recharging energy buffers.
While Chinchilla writes and reads from NVM to cross these periods, approximate intermittent computing can capture newer samples, as it already concluded processing the previous ones.


\section{Generality}
\label{sec:discussion}

The applicability of approximate intermittent computing extends beyond the case of human activity recognition or the use of specific machine learning techniques.
Here we summarize the design, implementation, and evaluation of a sharply different application.
Further details, including an extensive evaluation, are available~\cite{techrep2}.
We conclude by discussing the limitations of our approach.

\subsection{Application}

Embedded image processing enables applications such as smart parking, preventive maintenance, and pervasive surveillance~\cite{pedre2016accelerating}.
To accomplish the application tasks, image processing techniques such as corner detection are used to extract key features from the picture.
In a smart parking application, for example, corner information may be used to determine whether a spot is occupied.
The results of processing pictures at run-time are compared to the results obtained for a set of reference pictures, for example, showing the empty parking spot, to determine the final result.

Applications employing embedded image processing often exhibit the two characteristics discussed in the Introduction.
Current occupancies of parking spots must reach the end user as rapidly as possible; for example, to update public displays showing the number of available spots.
Occupancy information is as valid as it is most current; updating the information displays with stale information is essentially of no use.
Most importantly, corner detection only offers probabilistic guarantees in the first place, as most image processing techniques do.
The processing is already robust to data errors, for example, due to distortions in the pictures, and is therefore amenable to further approximation.

Embedded image processing using energy-harvesting devices enables zero-maintenance deployments, yet is extremely challenging.
The major issue is data size; even the simplest camera easily generates 25\kb of data for a single image capture~\cite{naderiparizi2015wispcam}.
Processing this amount of data for every frame may be prohibitive on energy-harvesting devices, requiring either very large energy buffers that require long times to recharge, increase footprints, and suffer from high leakage currents, or the frequent use of persistent state to stretch the processing across multiple execution rounds.

\subsection{Data Processing}

Similar to many image processing techniques, corner detection is implemented by applying forms of iterative processing.
We apply \emph{loop perforation}~\cite{han2013approximate} to create the knob that approximate intermittent computing requires to trade accuracy for energy consumption.
Loop perforation skips a certain fraction of the loop iterations to save resources.
Although application-specific policies to determine \emph{what} iterations to skip exist~\cite{han2013approximate}, the choice is most often random.
By properly tuning the fraction of loop iterations not executed as a function of available energy, loop perforation allows us to conclude processing before the first power failure, and hence to spare the need of persistent state and energy-hungry operations on NVM.

\begin{figure}[tb!]
	\centering
	\footnotesize
	\begin{tabular}{|p{0.14\textwidth}|c|c|}
		\hline
		 & \textbf{Anytime SVM}& \textbf{Loop perforation} \\
		\hline
		\hline
		 \emph{Approximation knob} & Number of features & Loop iterations\\
		\hline
		\emph{Energy estimation} & Single feature & Single loop iteration\\
		\hline    
		\emph{Output parameter} & Activity classification & Number/position of corners\\
		\hline       
	\end{tabular}
        \vspace{-2mm}
	\caption{Relation between key concepts of approximate intermittent computing in the applications we consider.}
	\label{tab:parallel}
        \vspace{-2mm}
\end{figure}

The key concepts emerging from the application of approximate intermittent computing to human activity recognition return here, as summarized in \figref{tab:parallel}.
The parameter that determines the level of approximation, and consequently the energy saving, is here the number of loop iterations not executed, similar to the features that the anytime support vector machines do not consider.
The extent of data processing as a function of available energy is here obtained by estimating, using the same tools as before~\cite{epic}, the energy consumed by the single loop iteration; likewise, this information is the energy to process the single feature in human activity recognition.
The output is represented by the number and position of detected corners, similar to the human activity classification obtained earlier.
Based on this information, we quantify the accuracy achieved; as there is no ``ground truth'' here, accuracy is defined relative to an execution that does not skip any loop iteration. 

\subsection{Evaluation}

We describe first our prototype implementation, along with the metrics and baselines we consider, and summarize the results next. 

\fakepar{Prototype} We create a hardware/software prototype based on a TI Launchpad equipped with the same MSP430-FR5659 MCU used earlier.
We use the on-board FRAM to store the test pictures we use for evaluation and the output of image processing.
We manually retrieve the latter from FRAM at the end of every experiment.
The energy cost for these operations is factored out.
We use the same 1470\uf capacitor and charging circuit as in \secref{sec:system}. 

\begin{figure}[!tb]
  \centering
   \includegraphics[width=0.8\columnwidth]{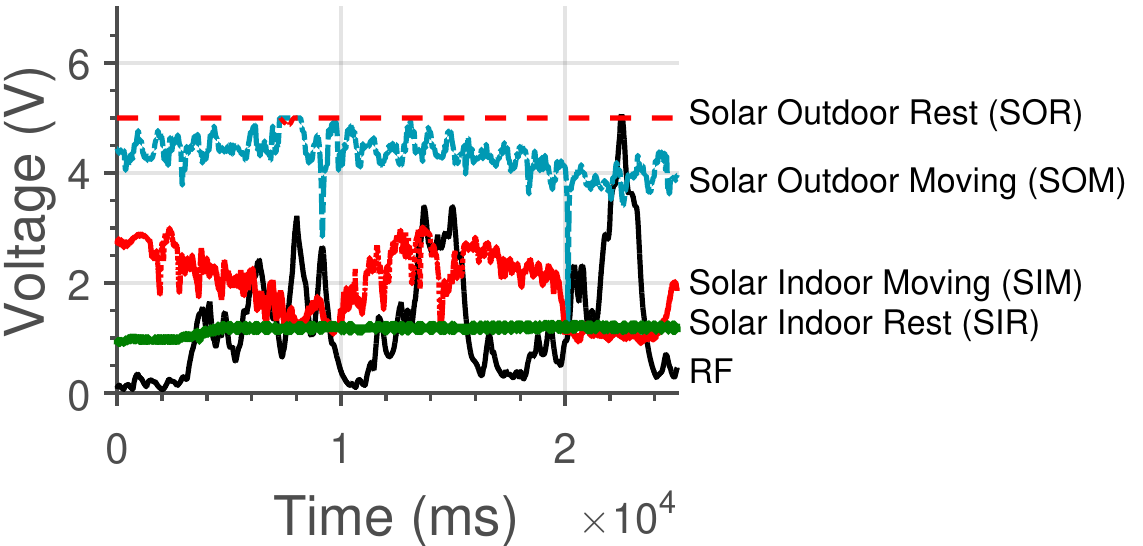}
   \vspace{-2mm}
    \caption{Energy traces used to power an embedded image processing pipeline. \capt{SOM is most stable and has highest energy content. RF is most variable and with least energy content.}}
   \label{fig:realworldtraces}
  \end{figure}

We supply energy using a Renesas digital power supply driven by an RL78/I1A controller, based on five energy traces obtained from diverse sources and in different settings.
\figref{fig:realworldtraces} shows an excerpt, plotting the instantaneous voltage reading over time.
The RF trace is from Mementos~\cite{mementos}, recorded using a WISP device~\cite{sample2008design}.
The other four traces are from EPIC~\cite{epic}  and are recorded using a mono-crystalline solar panel attached to an Arduino Uno in settings including outdoor mobile (SOM), indoor mobile (SIM), outdoor static (SOR), and indoor static (SIR). 
Similar to Ekho~\cite{EKHO}, this setup allows us to replicate the exact V-I curve the device would experience if attached to the actual energy harvester while considering the equivalent resistance offered by the device, and yet retain repeatability across experimental settings.

  \begin{figure}[tb]
    \centering
   \vspace{-2mm}
    \subfigcapskip 0pt
  \subfigure[Rectangle.]{
    \label{fig:square}
   \includegraphics[width=.99\linewidth]{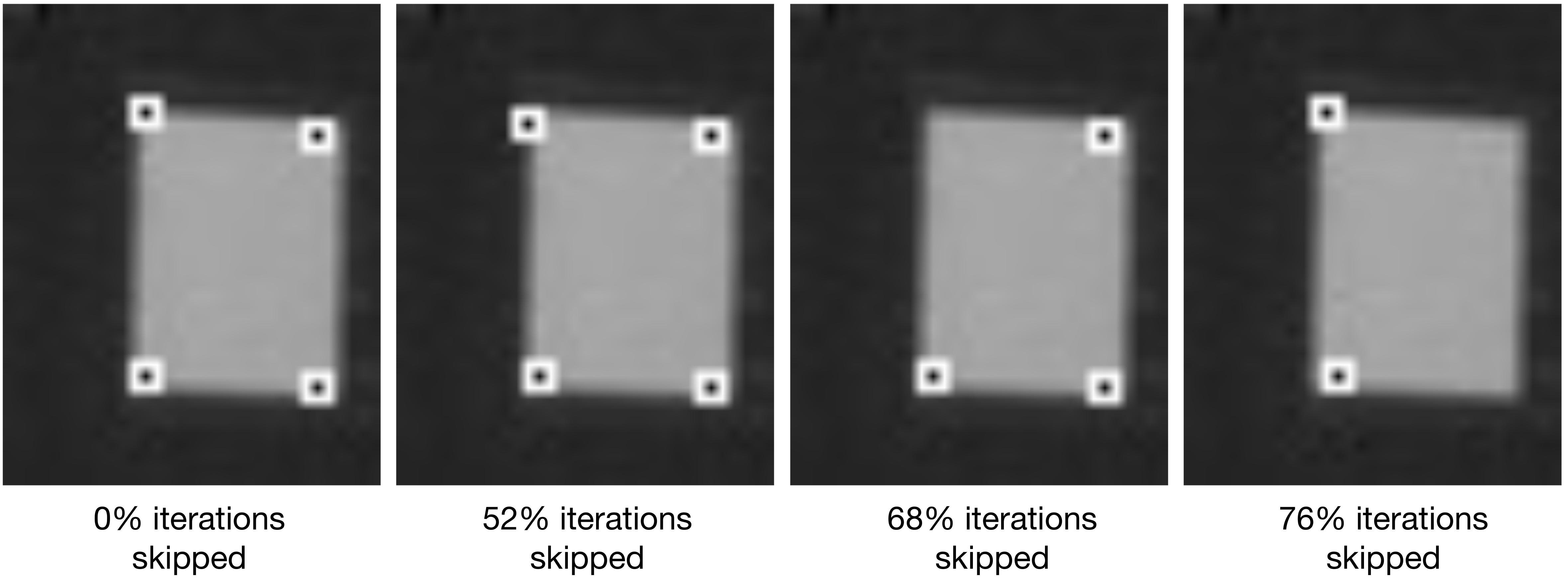}
  }
  \subfigure[Pen.]{
    \label{fig:pen}
   \includegraphics[width=.99\linewidth]{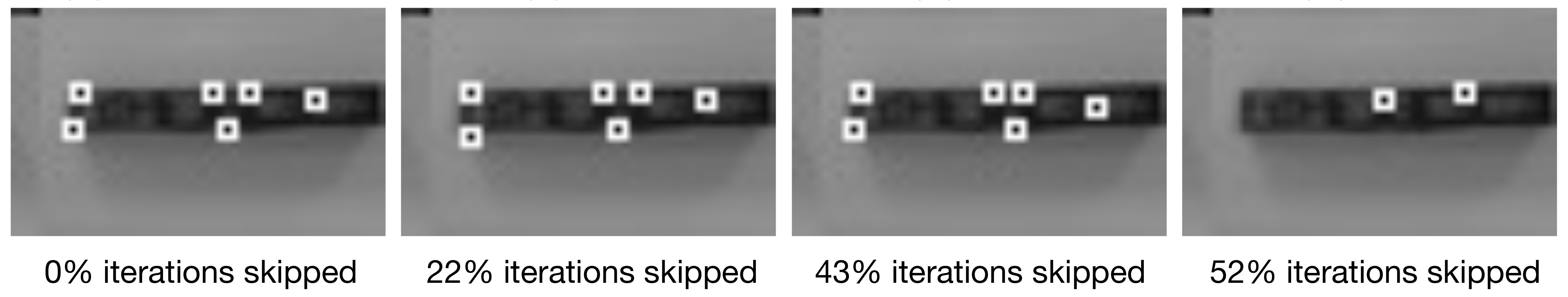}
  }
  \subfigure[Car.]{
    \label{fig:car}
   \includegraphics[angle=90,width=.99\linewidth]{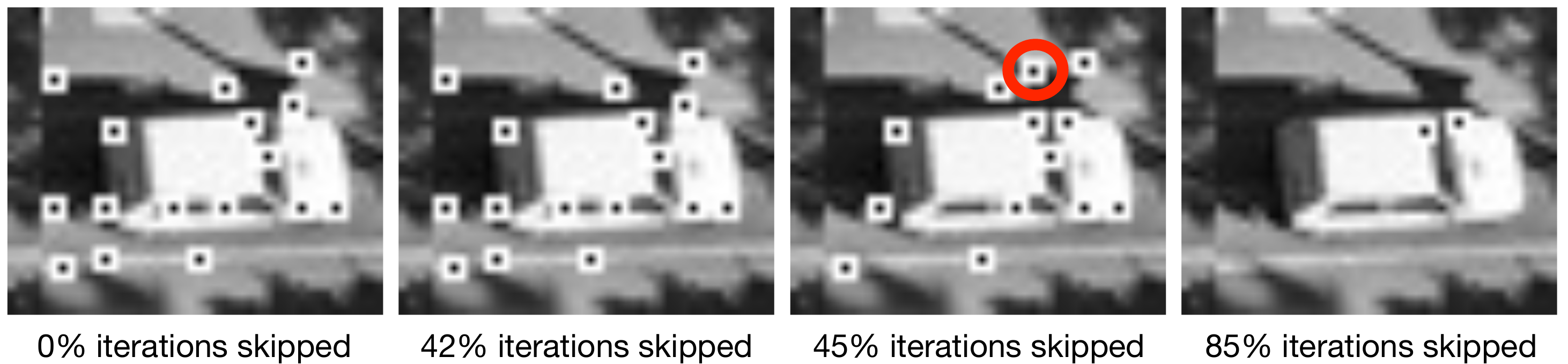}
 }
   \vspace{-3mm}
   \caption{Representative examples of the output of corner detection, depending on the fraction of loop iterations not executed. \capt{For complex pictures, up to 42\% of the loop iterations may be skipped without severely impacting the quality of the result.}}
     \vspace{-3mm}
  \label{fig:pics}
\end{figure}

Whenever the device wakes up with new energy, it randomly loads one of the test pictures and performs corner detection.
If energy is left at the end of the processing, the MCU switches to the lowest power mode that allows a 30\secs timer to eventually trigger another round of image processing.
The approximate intermittent computing implementation works the same as \greedy, skipping a number of loop iterations that allows the system to output the result just before energy is exhausted. 
We compare this again with a continuous execution and Chinchilla. 
Note that the absence of an actual camera module does not impact the relative performance figures, as the cost of image acquisition is paid no matter what is the implementation of data processing.

The metrics we consider next are the same as in \secref{sec:eval}, but measured according to the specific features of the application at hand.

\fakepar{Results} \figref{fig:pics} graphically shows representative examples of the outputs we obtain, as a function of the loop iterations skipped.
Note that the latter quantity here is precisely proportional to the share of saved energy. 
The picture in \figref{fig:square} is a simple test.
Approximate intermittent computing may skip up to more than half of the loop iterations, and yet return corner information that are equal in number, and very similar in positioning, compared to an execution that skips no iteration.
With more complex pictures, as in \figref{fig:pen} and \figref{fig:car}, this observation applies up to situations where more than 42\% of the loop iterations are not executed.
Beyond that, the overall number of detected corners reduces and spurious detections also appear, as indicated by the red circle in \figref{fig:car}.

  \begin{figure}[tb]
  \centering
  \includegraphics[angle=270,width=.8\linewidth]{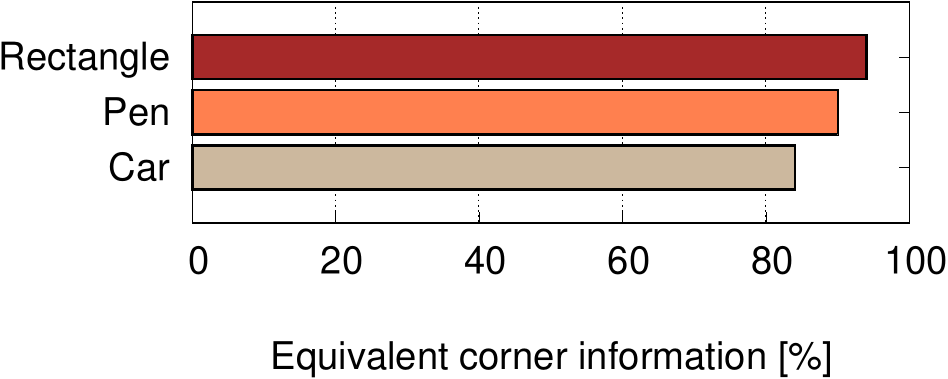}
    \vspace{-2mm}
    \caption{Quantifying the accuracy of approximate intermittent computing, compared to a continuous execution. \capt{Based on the number and position of detected corners, approximate intermittent computing returns an output that is equivalent to a continuous execution in at least 84\% of the cases.}}
        \vspace{-3mm}
\label{fig:picsQuantity}
\end{figure}

  \begin{figure}[tb]
  \centering
  \includegraphics[angle=270,width=.8\linewidth]{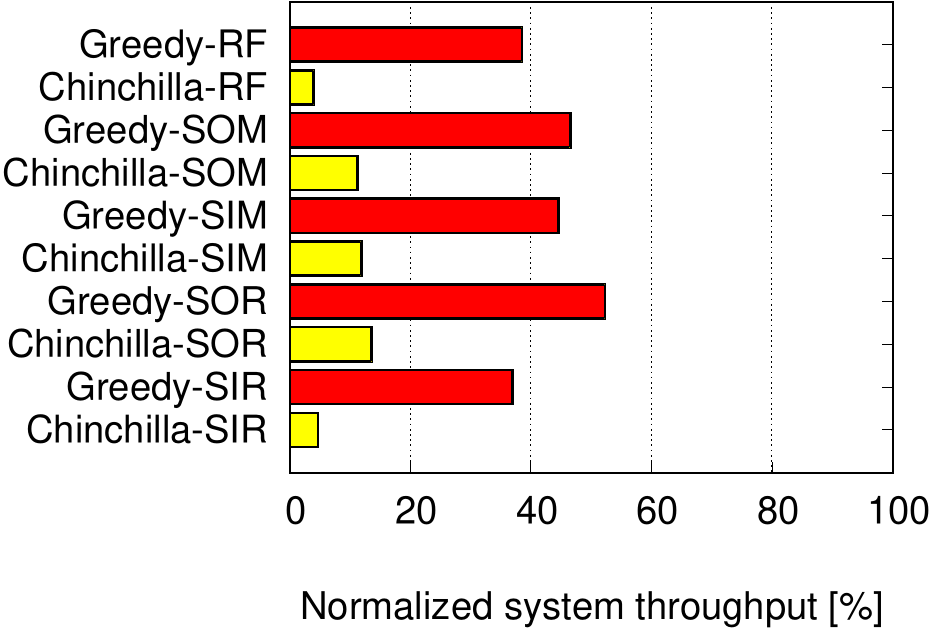}
    \vspace{-2mm}
  \caption{System throughput of corner detection normalized to that of a continuous execution. \capt{Because of the higher energy efficiency, traces with higher energy content amplify the gains of approximate intermittent computing.}}
    \vspace{-2mm}
  \label{fig:intermittentThroughputPics}
\end{figure}

\begin{figure*}[tb]
  \centering
  \includegraphics[width=\linewidth]{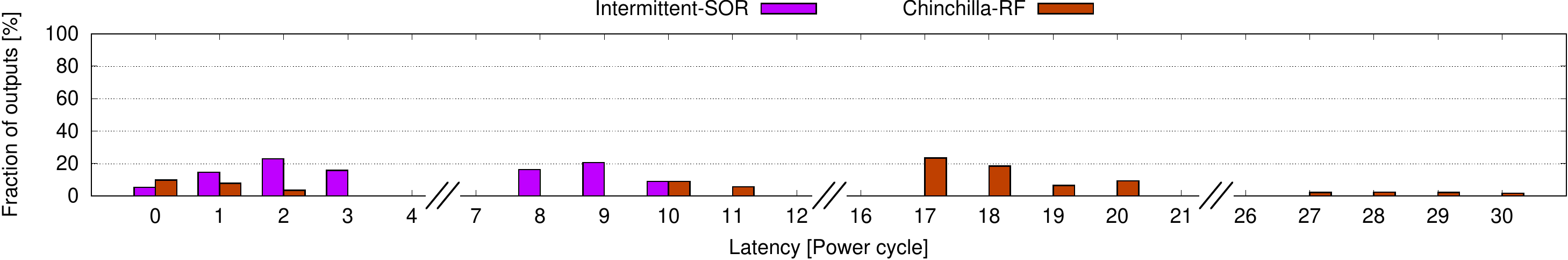}
      \vspace{-5mm}
  \caption{Distribution of the latency to produce the output of corner detection, measured in number of power cycles between when the picture is loaded until the output is produced. \capt{In situations of energy abundance, Chinchilla concludes the processing in at most ten power cycles, whereas highly dynamic energy traces cause higher latency.}}
    \vspace{-2mm}
  \label{fig:latencyIntermittentPics}
\end{figure*}

\figref{fig:picsQuantity} quantifies the average accuracy of approximate intermittent computing across all energy traces by showing the fraction of pictures whose corner information are equivalent to those obtained by a continuous execution that skips no loop iterations.
We define equivalence as the same number of corners appearing in the output, and their position in the approximate intermittent execution to be closer to the position of the same corner in the continuous execution than to any other corner.
The latter condition ensures that a corner may not be confused with a different one.
Based on this, for example, the leftmost three pictures in \figref{fig:pen} represent an equivalent output.
Depending on the complexity of the picture, approximate intermittent computing returns an equivalent output as a continuous execution in \emph{at least} 84\% of the cases. 

As seen in \secref{sec:eval}, approximate intermittent computing trades a loss in accuracy for the ability to produce the result in the same power cycle.
This reduces latency to return the result, enabling higher throughput.
\figref{fig:intermittentThroughputPics} shows the performance in the latter metric, normalized to that of a continuous execution.
The approximate intermittent computing implementation constantly outperforms the one using Chinchilla.
Generally, traces that are richer in energy correspond to larger improvements for approximate intermittent computing.
This is because it uses energy more efficiently than Chinchilla, where a significant fraction of that is spent handling persistent state and thus subtracted from application processing.

Interestingly, the better use of energy in approximate intermittent computing is visible in the time dynamics as well.
This aspect is apparent as one observes that the performance of approximate intermittent computing in \figref{fig:intermittentThroughputPics} is very similar for the RF and SIR traces: these two are very different in time, yet provide roughly the same total amount of energy to the system.
We make  better use of the energy available, no matter when it becomes available, by tuning the target accuracy, based on current conditions.
In contrast, Chinchilla suffers from the rapid dynamics in the RF trace.

\figref{fig:latencyIntermittentPics} concludes our analysis showing the latency to produce the result with two example traces.
Approximate intermittent computing is not shown here, as the result is produced within the same power cycle regardless of the energy trace.
The trends in \figref{fig:latencyIntermittent} match those in \figref{fig:latencyEmulation} and \figref{fig:latencyIntermittent}, showing how Chinchilla stretches the processing over many power cycles, as a function of the energy patterns.
In situations of energy abundance, as with the SOR trace, processing can conclude in at most ten power cycles.
The energy scarcity and rapid fluctuations of the RF trace, conversely, spread the processing over more power cycles.

\subsection{Limitations}

Scenarios where the data pipelines are amenable to approximation are, in principle, candidates for approximate intermittent computing.
The aforementioned examples of smart health~\cite{solanas2014smart}, ambient intelligence~\cite{cook2009ambient}, and environment monitoring~\cite{othman2012wireless} extend the range of applications we concretely demonstrate here.

The efficiency of the resulting system, however, is a function of how we can accurately estimate the energy consumption as a function of the level of approximation.
This is required to determine how far can the system go applying approximation.
Energy estimation tools for low-power embedded computing exist aplenty~\cite{konstantakos2008energy}, along with versions that are specific to intermittent computing~\cite{epic,cleancut,edb,SIREN}.
These tools are necessary companions to approximate intermittent computing.
Most of them work off-line, as the run-time overhead of energy estimation may be prohibitive.
Whenever these tools cannot provide precise estimates, for example, because executions are highly dependent on run-time information, approximate intermittent computing may be difficult to apply. 

By its own nature, approximate intermittent computing only offers probabilistic guarantees on correctness.
Many of the inaccurate results returned in our application prototypes, nonetheless, may be corrected through some form of post-processing, as they are often represented by single outliers in long sequences of accurate outputs.
This would further improve the accuracy of the system as a whole.
Despite this possibility, whenever applications require exact results or depend on the absolute precision of data, approximate intermittent computing is simply not applicable.
This is the case, for example, when sensor devices are used to drive closed-loop control systems, where accuracy is key to achieve stable behaviors~\cite{ploplys2004closed}.


\section{Conclusion}
\label{sec:end}

We presented the concept of approximate intermittent computing and demonstrated its application in two diverse application cases.
Regular intermittent computing retains the equivalence to continuous executions by using persistent state on NVM to cross periods of energy unavailability.
In contrast, we showed how a moderate loss in the accuracy of the output provides huge gains in terms of energy consumption in that, if properly tuned, the system can finish processing prior to the first power failures.
This makes it possible for the system to spare the need to maintain persistent state on the energy-hungry NVM, and thus allows the energy budget to be spent \emph{entirely} for useful application processing.
As a result, outputs are returned to the end user within the same power cycle compared to when the input sensor data is gathered.
System throughput increases consequently, as the system is ready to process new inputs as soon as it restarts after a power failure.
We showed, for example, that in human activity recognition approximate intermittent computing provides a $7x$ improvement in system throughput compared to regular intermittent computing.
It also retains an average 83\% accuracy compared to ground truth, in a setting where the best attainable accuracy is 88\%.
Using imagine processing techniques for corner detection, we achieve a $5x$ improvement in system throughput compared to regular intermittent computing, while retaining equivalence in the number and position of detected corners in at least 84\% of the cases, compared to a continuous execution.



\newpage
\bibliographystyle{ACM-Reference-Format}
\bibliography{biblio}

\end{document}